\documentclass[letter]{aa}  

\usepackage{graphicx}
\usepackage{txfonts}
\usepackage{gensymb}
\usepackage{hyperref}

\begin{document}

   \title{A high HDO/H$_2$O ratio in the Class~I protostar L1551 IRS5\thanks{This work is based on observations carried out under project numbers W18AO and S16AE with the IRAM NOEMA Interferometer. IRAM is supported by INSU/CNRS (France), MPG (Germany), and IGN (Spain).}}
   \titlerunning{A high HDO/H$_2$O ratio in the Class~I protostar L1551 IRS5}

   \author{A. Andreu \inst{1}  
          \and
          A. Coutens \inst{1}
          \and
         F. Cruz-S\'aenz de Miera \inst{1,2,3}
         \and
         N. Houry \inst{1} 
         \and
         J. K. J\o rgensen \inst{4}
         \and
         A. K\'osp\'al \inst{2,3,5,6}
         \and
         D. Harsono \inst{7}
        }

   \institute{Institut de Recherche en Astrophysique et Plan\'etologie (IRAP), Universit\'e de Toulouse, UT3-PS, CNRS, CNES, 9 av. du Colonel Roche, 31028 Toulouse Cedex 4, France \\
              \email{audrey.andreu@irap.omp.eu, audrey.coutens@irap.omp.eu}
         \and 
         Konkoly Observatory, Research Centre for Astronomy and Earth Sciences, E\"otv\"os Lor\'and Research Network (ELKH), Konkoly-Thege Mikl\'os \'ut 15-17, 1121 Budapest, Hungary
         \and
         CSFK, MTA Centre of Excellence, Budapest, Konkoly Thege Mikl\'os \'ut 15-17., H-1121, Hungary
             \and
         Centre for Star and Planet Formation, Niels Bohr Institute \& Natural History Museum of Denmark, University of Copenhagen, \O ster Voldgade 5-7, DK-1350 Copenhagen K, Denmark
          \and
             Max-Planck-Institut f\"ur Astronomie, K\"onigstuhl 17, D-69117 Heidelberg, Germany
             \and
             E\"otv\"os Lor\'and University, Department of Astronomy, P\'azm\'any P\'eter s\'et\'any 1/A, 1117 Budapest, Hungary
            \and
            Institute of Astronomy, Department of Physics, National Tsing Hua University, Hsinchu, Taiwan}

   \date{Received xxx; accepted xxx}

  \abstract
  {
  Water is a very abundant molecule in star-forming regions. Its deuterium fractionation provides an important tool for understanding its formation and evolution during the star and planet formation processes. While the HDO/H$_2$O abundance ratio has been determined toward several young Class~0 protostars and comets, the number of studies toward Class~I protostars is limited.}
  {Our aim is to study the water deuteration toward the Class~I protostellar binary L1551 IRS5 and to investigate the effect of evolutionary stage and environment on variations in the water D/H ratio.} 
  {Observations were carried out toward L1551 IRS5 using the NOrthern Extended Millimeter Array (NOEMA) interferometer. The HDO 3$_{1,2}$--2$_{2,1}$ transition at 225.9 GHz and the H$_2^{18}$O 3$_{1,3}$--2$_{2,0}$ transition at 203.4 GHz were covered with a spatial resolution of 0.5$\arcsec \times$ 0.8$\arcsec$, while the HDO 4$_{2,2}$--4$_{2,3}$ transition at 143.7 GHz was observed with a resolution of 2.0$\arcsec$ $\times$ 2.5$\arcsec$. We constrained the water D/H ratio using  both local thermodynamic equilibrium (LTE) and non-LTE models.}
 {The three transitions are detected. The line profiles display two peaks, one at $\sim$6 km\,s$^{-1}$ and one at $\sim$9 km\,s$^{-1}$. We derive an HDO/H$_2$O ratio of (2.1 $\pm$ 0.8) $\times$ 10$^{-3}$ for the redshifted component
  and a lower limit of $>$ 0.3 $\times$ 10$^{-3}$ for the blueshifted component. This lower limit is   
  due to the blending of the blueshifted H$_2^{18}$O component with redshifted CH$_3$OCH$_3$ emission.} 
  {The HDO/H$_2$O ratio in L1551~IRS5 is similar to the values in Class~0 isolated sources and in the disk of the Class~I protostar V883 Ori, while it is significantly higher than in the previously studied clustered Class~0 sources and the comets. This result suggests that the chemistry of protostars that belong to molecular clouds with relatively low source densities, such as L1551, share more similarities with the isolated sources than the protostars of very dense clusters. If the HDO/H$_{2}$O ratios in Class~0 protostars with few sources around are comparable to those found to date in isolated Class~0 objects, it would mean that there is little water reprocessing from the Class~0 to Class~I protostellar stage.}


   \maketitle
%

\section{Introduction}

Water is one of the most abundant molecules in the interstellar medium. Its role is significant in the star formation process as it cools down the warm regions and boosts the gravitational collapse.  It has been detected in multiple environments, such as protostars \citep[e.g.,][]{vanDishoeck2021}, protoplanetary disks \citep[e.g.,][]{Hogerheijde2011}, comets \citep[e.g.,][]{Mumma1986,Davies1997}, and asteroids \citep[e.g.,][]{Rivkin2010}. 
Deuteration is extremely sensitive to density and temperature conditions \citep{Ceccarelli2014}, and higher D/H ratios are expected  in particular when water forms in cold and dense conditions \citep[e.g.,][]{Coutens2012, Jensen2019}. The comparison of the D/H abundance ratio of the terrestrial oceans (Vienna Standard Mean Ocean Water (VSMOW) $\sim$ 1.557 $\times$ 10$^{-4}$, \citealt{deLaeter2003}) with comets and asteroids (a few 10$^{-4}$) suggests that these primitive objects could have delivered part of the water found on Earth \citep[e.g.,][]{Hartogh2011,Altwegg2015,TrigoRodriguez}. 
Chemical models of water deuteration in protoplanetary disks also show that the fractionation is not efficient enough to reproduce the terrestrial water D/H ratio without considering an inheritance from the parental molecular cloud \citep{Cleeves2014}. 
Studying the water deuteration in protostars can consequently help understand the origin of water in disks and planets \citep{Furuya2016}.

During the last decade the water deuterium fractionation ratio\footnote{The D/H ratio of water is equal to 1/2 $\times$ HDO/H$_2$O due to statistics.} has been measured in the warm inner regions of several Class~0 protostars, thanks to interferometric observations with the Atacama Large Millimeter/submillimeter Array (ALMA), the Submillimeter Array, and the Plateau de Bure Interferometer \citep[][and references thereafter]{Jorgensen2010, Taquet2013, Coutens2014}. 
In the clustered sources (IRAS 16293$-$2422, NGC1333 IRAS2A, IRAS4A$-$NW, and IRAS4B) the HDO/H$_2$O ratio was found to range between 6 $\times$ 10$^{-4}$ and 19 $\times$ 10$^{-4}$ \citep{Persson2013,Persson2014}. 
However, for the isolated sources (L483, B335, and BHR71$-$IRS1) \citet{Jensen2019} found a HDO/H$_2$O ratio of (1.7 - 2.2) $\times$ 10$^{-3}$, which is  a factor of $\sim$2--4 higher than for the clustered sources. 
This difference was proposed to be the consequence of  different durations or temperatures in the prestellar phase \citep{Jensen2021}.

To reconstruct the full history of water during the star formation process and bridge the gap between the Class 0 protostellar stage and the formation of planets, comets, and asteroids, studies of more evolved protostars are  needed. Even if the detection of water is usually difficult in such sources \citep{Harsono2020}, there are a few Class I objects in which water can be detected, for example after a strong increase in the accretion rate from the disk onto the protostar \citep[e.g., FUor or EXor sources,][]{Audard2014,Fischer2022}. Such an event leads to a luminosity burst that warms up the disk and the envelope in which the icy grain mantles can sublimate \citep{Cieza2016}. However, the studies were  limited to only two sources. A first study by \citet{Codella2016} showed the detection of HDO toward the EXor-like source SVS13-A  with the IRAM 30 m telescope. The presence of high-energy transitions suggests that they come from the warm inner regions even though emission from large-scale shocks cannot be excluded. From the nondetection of the H$_2^{18}$O transition at 203.4 GHz, they estimated a lower limit of the HDO/H$_2$O ratio of $\gtrsim$ 10$^{-3}$.
More recently, a HDO/H$_2$O ratio of (2.26 $\pm$ 0.63) $\times$ 10$^{-3}$ was obtained with ALMA in the disk of the FUor protostar V883 Ori \citep{Tobin2023}. The similarity of the water deuteration in isolated Class~0 protostars supports the inheritance of water from the envelope to the disk.

Here we report the interferometric determination of the HDO/H$_2$O ratio in another Class I source, L1551 IRS5 \citep{Chen1995}. This protostellar binary, located in the Taurus molecular cloud (d $\sim$ 155 pc, \citealt{Zucker2019}), is classified as a FUor-like object by \citet{Connelley2018} based on its near-infrared spectra \citep{Mundt1985,Carr1987}. The detection of complex organic molecules in its inner regions suggests high-temperature conditions that would be sufficient to thermally desorb the molecules frozen on grains, similarly to hot corinos \citep[Cruz-S\'aenz de Miera, in prep.]{Bianchi2020}.

\begin{table*}[t!]
\caption{Detected HDO and H$_2^{18}$O transitions toward the L1551 IRS5 source.}  
\begin{center}
\begin{tabular}{lcccccccccc}
\hline
\hline
Molecule & Transition & Frequency  & $E_{\rm up}$ & $A_{\rm ij}$  & $g_{\rm up}$ & Beam size & PA & rms & $d\varv$ & Project \\
 &  & [MHz]  & [K] & [s$^{-1}$] & & [$\arcsec$ $\times$ $\arcsec$] & [\degree] & [mJy beam$^{-1}$]  & [km\,s$^{-1}$] & \\
\hline
HDO & 3$_{1,2}$ - 2$_{2,1}$ & 225896.72 & 168 & 1.3 $\times$ 10$^{-5}$ & 7 & 0.76 $\times$ 0.51 & 21.7 & 4.2 & 0.5 & W18AO \\
H$_2^{18}$O & 3$_{1,3}$ - 2$_{2,0}$ & 203407.52 & 204 & 4.8 $\times$ 10$^{-6}$ & 7 & 0.83 $\times$ 0.57 & 24.0 & 3.7 & 0.5 & W18AO \\
HDO & 4$_{2,2}$ - 4$_{2,3}$ & 143727.21 & 319 & 2.8 $\times$ 10$^{-6}$ & 9 & 2.51 $\times$ 2.05 & 37.1 & 2.1 & 4.0 & S16AE \\
\hline
\end{tabular}
\end{center}
\label{tab:target-lines}
\end{table*}

\section{Observations}
\label{sect_obs}

Observations of two HDO transitions and one H$_2^{18}$O transition were carried out toward L1551 IRS5 with the Northern Extended Millimeter Array (NOEMA) through the projects W18AO and S16AE (see Table \ref{tab:target-lines}). The HDO 3$_{1,2}$--2$_{2,1}$ and H$_2^{18}$O 3$_{1,3}$--2$_{2,0}$ transitions at 225.9 GHz and 203.4 GHz, respectively, were observed simultaneously in the A configuration on 15 and 17 February 2019 and in the C configuration on 27 February 2019. 
High spectral resolution windows with channel spacing of 62.5 kHz (i.e., 0.08 -- 0.09~km\,s$^{-1}$) were used to cover the two transitions. We resampled the data with a resolution $d\varv$ of 0.5 km\,s$^{-1}$ to increase the signal-to-noise ratio. 
The HDO 4$_{2,2}$--4$_{2,3}$ transition at 143.7 GHz was covered with a poor resolution of 1.95 MHz  (i.e., 4 km\,s$^{-1}$) in other NOEMA observations carried out on 25 August and 31 August 2016 in the D configuration. 
All the data were calibrated and imaged using the CLIC and MAPPING packages of the GILDAS\footnote{\url{http://www.iram.fr/IRAMFR/GILDAS/}} software. 
The list of calibrators for each observation is in Table \ref{tab:calibrators}. The continuum was subtracted before the Fourier transform was applied to  the line data. The data were imaged with the natural weighting mode. 
The synthesized beam sizes and rms obtained for each transition are listed in Table \ref{tab:target-lines}. The two components, N and S, of L1551 IRS5 are not fully resolved in our observations as they are only separated by 0.36$\arcsec$ (i.e., $\sim$56 AU, \citealt{Cruz2019}).

\section{Results}
\label{sect_results}

\begin{figure*}[ht!]
    \centering
    \resizebox{\hsize}{!}{
    \includegraphics{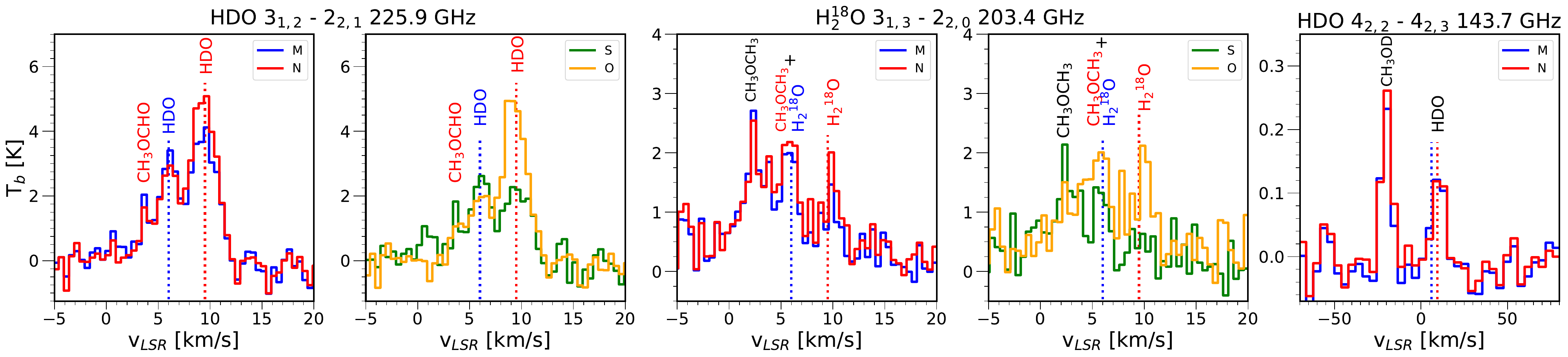}}
    \caption{
    Comparison of the spectra of the water isotopolog transitions. The green, blue, red, and orange solid lines represent the spectra extracted at the position S, M, N, and O, respectively.  
    The dotted lines show velocities of 6.0 km\,s$^{-1}$  (blue) and 9.5 km\,s$^{-1}$ (red). Considering the synthetic beam sizes and the separation of the N and S components, the extracted spectra are not entirely independent from each other. For the HDO line at 143.7 GHz, the spatial resolution is too low to  distinguish O from N and M from S. The x-axis scale is larger for this transition.}
    \label{fig:super-line}
\end{figure*}

\begin{figure}
    \resizebox{\hsize}{!}{
    \includegraphics{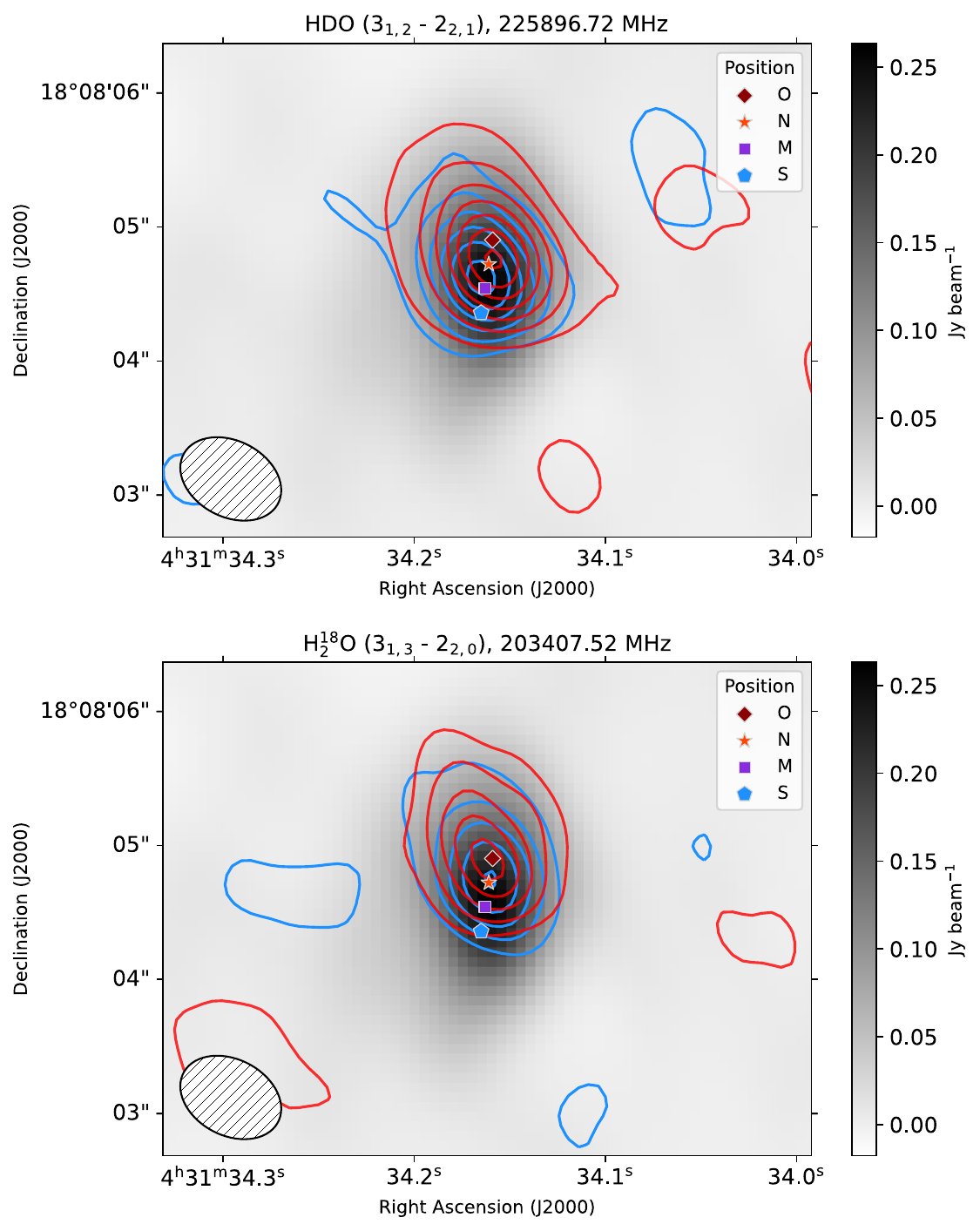}}
    \caption{
    Integrated emission maps of the HDO line at 225.9 GHz (top) and the H$_2^{18}$O line at 203.4 GHz (bottom). The blue and red contours correspond to the integrated flux between 4.5 and 7.5 km\,s$^{-1}$ and 7.5 and 12 km\,s$^{-1}$, respectively. The levels of contours start at 3$\sigma$ with a step of 3$\sigma$ for all except for the HDO contours in red (7.5--12 km\,s$^{-1}$) which are spaced by 5$\sigma$. Dust continuum is indicated in grayscale. The beam sizes are shown in the bottom left corner.     }
    \label{fig:MAPS}
\end{figure}

In a first step, we visualized the datacubes of the HDO and H$_2^{18}$O lines with the Aladin\footnote{\url{https://aladin.u-strasbg.fr/AladinDesktop/}} and CASSIS$\footnote{CASSIS (\url{http://cassis.irap.omp.eu/}) has been developed by IRAP-UPS/CNRS.}$ softwares \citep{Bonnarel2000,Glorian2021}.
We extracted the spectra at the positions of the northern source ($\alpha_{\rm J2000}$=04$^{\rm h}$31$^{\rm m}$34$\fs$161, $\delta_{\rm J2000}$=+18$\degr$08$\arcmin$04.72$\arcsec$) and the southern source ($\alpha_{\rm J2000}$=04$^{\rm h}$31$^{\rm m}$34$\fs$165, $\delta_{\rm J2000}$=+18$\degr$08$\arcmin$04.359$\arcsec$, \citealt{Takakuwa2020}). 
To study variations in the line intensities with the position, the spectra were also extracted for a position between the two sources, at an equal distance from them   ($\alpha_{\rm J2000}$=04$^{\rm h}$31$^{\rm m}$34$\fs$163, $\delta_{\rm J2000}$=+18$\degr$08$\arcmin$04.550$\arcsec$), that we  call M. Because of the asymmetry in emission maps with respect to M, we include a position at a similar distance from N in the opposite direction ($\alpha_{\rm J2000}$=04$^{\rm h}$31$^{\rm m}34\fs$159, $\delta_{\rm J2000}$=+18$\degr$08$\arcmin$04.890$\arcsec$) called O. 
A comparison of these spectra is shown in Figure \ref{fig:super-line}, and their locations are indicated in Figure \ref{fig:MAPS}.

Two velocity components are seen in the line profiles at 203.4 and 225.9 GHz, similarly to \citet{Bianchi2020} and \citet{Mercimek2022} for complex organic molecules. The component at $\sim$9 km\,s$^{-1}$ is brighter at the N and O positions than at the M and S positions. The HDO component at $\sim$6 km\,s$^{-1}$ is slightly brighter at the M position than the N and S positions, while it is less bright at the O position. The high-velocity peak   corresponds to the N source or farther to the north, in agreement with \citet{Bianchi2020}. 
The HDO component at 6 km\,s$^{-1}$ seems to arise closer to the M position than the S position based on the comparison of the observed line intensities.
The double peak created by these two velocity components could then possibly indicate rotation from a disk surrounding the N source. 
For the HDO line at 143.7 GHz, the spatial resolution is too low ($\sim$2.0-2.5$\arcsec$) compared to the binary separation (0.36$\arcsec$) to observe spectral differences at the four positions (see Fig. \ref{fig:super-line}).
The two velocity components are not observed because of the low spectral resolution of the data (4 km\,s$^{-1}$).

We used the CASSIS software to check possible blending with other molecules at velocities of both 6 km\,s$^{-1}$ and 9 km\,s$^{-1}$. The spectroscopic data come from the CDMS and JPL databases \citep{Pickett1998,Muller2005}. No blending is observed for the component at $\sim$ 9 km s$^{-1}$ either for HDO or H$_2^{18}$O. However, the H$_2^{18}$O component at $\sim$ 6 km s$^{-1}$ is blended with a CH$_3$OCH$_3$ 3$_{3,1,1}$ - 2$_{2,1,1}$ line emitted at $\sim$ 9 km\,s$^{-1}$.
A faint line of CH$_3$OCHO 6$_{6,0}$ - 5$_{5,0}$ (emission at 9 km s$^{-1}$) is observed in the wing of the HDO transition at 225.9 GHz. The HDO line at 143.7 GHz does not seem to be affected by blending.

Integrated emission maps of the HDO 225.9 GHz and H$_2^{18}$O 203.4 GHz lines are shown in Figure \ref{fig:MAPS}. The water emission is compact (less than 70 AU in diameter), as also seen for complex organic molecules \citep{Bianchi2020}. 
As expected, the HDO component at 9 km\,s$^{-1}$ (in red) peaks close to the N  position, while the one at 6 km\,s$^{-1}$ (in blue) peaks closer to the M position. 
A similar distribution is seen for the H$_2^{18}$O component at 9 km\,s$^{-1}$. However, the one at 6 km\,s$^{-1}$ peaks closer to the N position. This can be explained by the blending with the CH$_3$OCH$_3$ line, which is emitted at a velocity of 9 km\,s$^{-1}$. It consequently means that the CH$_3$OCH$_3$ emission contributes significantly to the flux observed at 6 km\,s$^{-1}$. The HDO and H$_2$$^{18}$O emission sizes were estimated with circular Gaussian fitting of the lines in the ($u$,$\varv$)-plane. We found approximate source sizes of 0.35$\arcsec$ and 0.45$\arcsec$ for the components at 6 and 9 km\,s$^{-1}$, respectively (see Table \ref{tab:uv-fits}).

To determine the water deuteration for the two velocity components, we used the HDO 225.9 GHz and H$_2^{18}$O 203.4 GHz lines. We focused our analysis on the spectra extracted at the position of the N source for the component at 9 km\,s$^{-1}$ and the spectra extracted at the position M (between N and S) for the component at 6 km\,s$^{-1}$. 
The line fluxes of the two components were collected by fitting the HDO and H$_2^{18}$O lines with double Gaussian profiles with the Levenberg-Marquardt fitter (see Figure \ref{fig:Gaussians}).
The results of the fits are presented in Table \ref{tab:gauss-param}. Adding a third Gaussian for the CH$_3$OCHO peak does not affect the HDO flux within the uncertainties. 
The Gaussian fluxes also do not differ  if we fix the same FWHM and $\varv_{\rm LSR}$ for the HDO and H$_2^{18}$O lines. Based on the Gaussian fits, we calculated the column densities of each isotopolog assuming local thermodynamic equilibrium (LTE),  following the formalism of \citet{Goldsmith1999} and \citet{Mangum2015}. 
Then we derived the HDO/H$_2$O ratios assuming a $^{16}$O/$^{18}$O ratio equal to 560 \citep{Wilson1994}. We considered a range of excitation temperatures, $T_{\rm ex}$, between 100 and 300 K, which are characteristic of hot corinos \citep[e.g.,][]{Jorgensen2018}.
Source sizes of 0.35$\arcsec$ and 0.45$\arcsec$ were assumed for M and N, respectively. The lines are optically thin ($\tau$ $\leq$ 0.23 for HDO and $\tau$ $\leq$ 0.15 for H$_2^{18}$O) in both cases. 
Figure \ref{fig:Tplotvar} shows the variation in the HDO/H$_2$O ratio as a function of the excitation temperature. The extreme values found for the HDO/H$_2$O ratio  are listed in Table \ref{tab:results}. The ratio varies between (1.6$\pm$0.6)\,$\times$\,10$^{-3}$ and (2.1$\pm$0.8)\,$\times$\,10$^{-3}$ for the component at 9 km\,s$^{-1}$. 
Only a lower limit of $>$0.3\,$\times$\,10$^{-3}$ can be derived for the component at 6 km\,s$^{-1}$
given the blending of the H$_2^{18}$O line with CH$_3$OCH$_3$. Using a different size does not significantly affect the HDO/H$_2$O ratio (see Appendix \ref{AppendixB1}). Non-LTE analysis also leads to similar results (see Appendix \ref{AppendixB2}).

As the fluxes of the two components at 6 and 9 km\,s$^{-1}$ are not disentangled in the HDO observation at 143.7 GHz, we could not use it to derive the HDO/H$_2$O ratios. However, it can help us constrain the excitation temperatures of HDO. The analysis described in Appendix \ref{AppendixB3} shows that a model with an excitation temperature for N equal to or lower than 220 K  cannot reproduce the flux of the 143.7 GHz line within 20\%. The HDO/H$_2$O ratio derived for $T_{\rm ex}$ = 300 K should consequently be considered more reliable than the value obtained for  $T_{\rm ex}$ = 100 K. 

\begin{table*}
\caption{Column densities and HDO/H$_2$O ratio in L1551~IRS5 for different excitation temperatures.}
\begin{center}

\begin{tabular}{lcccc}
\hline
\hline
 & \multicolumn{2}{c}{Component 9 km s$^{-1}$, position N}& \multicolumn{2}{c}{Component 6 km s$^{-1}$, position M} \\
 & $T_{\rm ex}$ = 100\,K & $T_{\rm ex}$ = 300\,K & $T_{\rm ex}$ = 100\,K & $T_{\rm ex}$ = 300\,K \\
\hline

\multicolumn{5}{c}{Column density [cm$^{-2}$]} \\
\hline
HDO & (7.5 $\pm$ 0.7) $\times$ 10$^{15}$ & (1.3 $\pm$ 0.2) $\times$ 10$^{16}$ & (4.8 $\pm$ 0.1) $\times$ 10$^{15}$ & (8.0 $\pm$ 1.7) $\times$ 10$^{15}$ \\

H$_2^{18}$O & (8.4 $\pm$ 2.5) $\times$ 10$^{15}$ & (1.1 $\pm$ 0.3) $\times$ 10$^{16}$ & $<$ (1.6 $\pm$ 0.3) $\times$ 10$^{16}$ & $<$ (2.1 $\pm$ 0.4) $\times$ 10$^{16}$ \\
\hline

\multicolumn{5}{c}{Isotopic ratio} \\
\hline
HDO/H$_2$O & (1.6 $\pm$ 0.6) $\times$ 10$^{-3}$ & (2.1 $\pm$ 0.8) $\times$ 10$^{-3}$ & $>$ (0.5 $\pm$ 0.2) $\times$ 10$^{-3}$ & $>$  (0.7 $\pm$ 0.3) $\times$ 10$^{-3}$ \\
\hline
\end{tabular}

\tablefoot{
    The source sizes are assumed to be 0.35$\arcsec$ and 0.45$\arcsec$ for M and N, respectively. The error bars are based on the Gaussian fit uncertainties. }
\end{center}
\label{tab:results}
\end{table*}

\section{Discussion} 
\label{sect_discu}

\begin{figure*}
    \centering  
    \includegraphics[width=17cm]{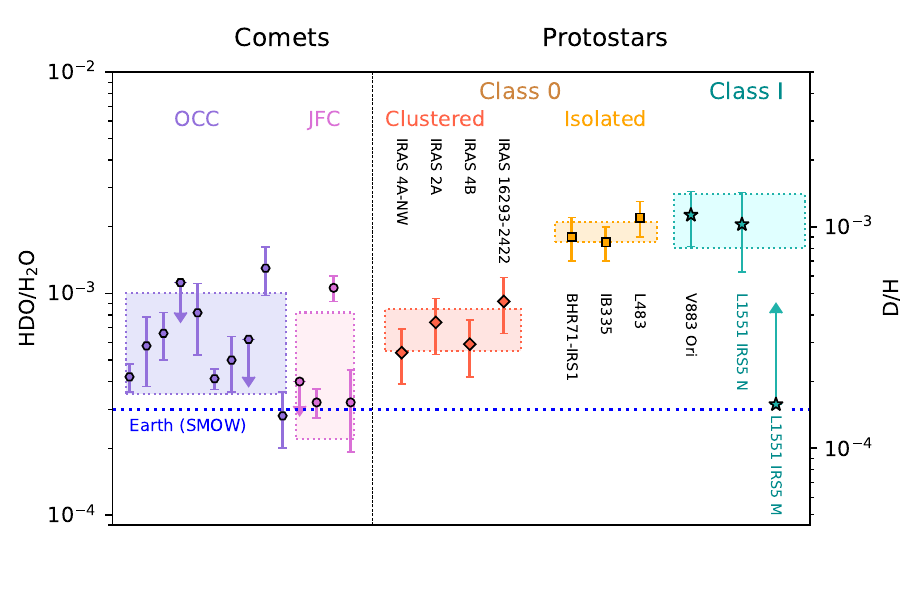} 
    \vspace{-0.6cm}
    \caption{Comparison of   HDO/H$_2$O ratios in comets and protostars. The values and references can be found in Table \ref{tab:Jensen-ratios}. The equivalent D/H ratio is indicated on the right axis. For L1551 IRS5 N ($\varv_{\rm LSR}$ $\sim$ 9 km\,s$^{-1}$), the ratio is plotted for an excitation temperature of 300 K. }
    \label{fig:plot-Jensen}
\end{figure*}

Figure \ref{fig:plot-Jensen} shows a comparison of the HDO/H$_2$O ratio of L1551~IRS5 with the values previously found in protostars and comets. 
For the component at 9 km\,s$^{-1}$, the HDO/H$_2$O ratio is similar to those measured in isolated Class~0 protostars \citep{Jensen2019} and in the disk of the Class~I V883 Ori \citep{Tobin2023}, and significantly higher than the previously studied clustered sources. 
At 6 km\,s$^{-1}$, the lower limit does not provide sufficient constraints to discriminate between the two categories. 

Water observed in the warm inner regions of protostellar objects results from the thermal desorption of the icy grain mantles. Deuteration takes place early in the star formation process when the temperature is low and the density is relatively high. As a result, the gas phase HDO/H$_2$O ratio observed in the warm regions of Class 0 objects should reflect those of the icy grain mantles. Consequently, the observed deuterium fractionation ratio at the Class~I stage should either be similar to the Class~0 stage or decrease if water is partially reprocessed in the hot corino. In the latter case the deuteration process would not be efficient enough to increase or maintain the overall deuteration level.
From the HDO/H$_2$O ratio found in the high-velocity component of L1551 IRS5 we can conclude that the HDO/H$_2$O ratio does not decrease and is still quite high, similarly to what is found for V883 Ori \citep{Tobin2023} and SVS13-A \citep{Codella2016}.

A second conclusion is that the chemistry of L1551 IRS5 appears to be closer to isolated sources than to clustered sources. 
The similarity with isolated sources can be surprising at first sight, as L1551~IRS5 is part of the Taurus molecular cloud. However, it is located in the LDN 1551 group in the southern part of the cloud \citep[e.g.,][]{Roccatagliata2020}. This group shows a lower density of sources than other regions in Taurus \citep{Gomez1993}. The spatial density of sources in the Taurus molecular cloud is also much lower (only a few tens  in $\sim$1 pc$^3$, \citealt{Gomez1993}) than in Ophiuchus, which contains IRAS~16293-2422, and the NGC1333 cloud in Perseus (a few 10$^2$--10$^3$, \citealt{Bontemps2001,Jorgensen2006}). 
As explained in \citet{Jensen2021}, the higher water deuteration  in isolated sources would be due to either a lower temperature or a longer prestellar phase compared to clustered sources. The environment around L1551 IRS5 could be colder than around IRAS16293 and the NGC1333 sources. The dense cores could also have more time to evolve before their collapse if L1551 IRS5 is on the edge of the cloud and if it is surrounded by fewer sources. As a consequence, the water deuteration in L1551 IRS5 would be closer to the Class 0 isolated sources.
A similar explanation was proposed for the Class I protostar V883 Ori, which belongs to the Orion molecular cloud, but is in a relatively isolated area \citep{Tobin2023}. The lower values found in comets and in the protostars IRAS~16293-2422, NGC1333 IRAS2A, IRAS4A, and IRAS4B suggest that our Solar System formed in a relatively dense clustered area. 
The similarity of the HDO/H$_2$O ratio between the binary L1551 IRS5 and the singly Class I object V883 Ori also seems to indicate that the binarity does not influence the water deuteration.  

In conclusion, this study highlights the importance of carrying out water deuteration measurements toward more sources, at different evolutionary stages, and in different environments in order to better understand the water formation and evolution during the formation process of solar-type stars.  Investigations of the D$_2$O/HDO ratio in Class I protostars would also be useful to check the similarity  with Class~0 sources.

\begin{acknowledgements}
The authors thank the IRAM staff, especially Jan Martin Winters and J\'er\'emie Boissier for their help with the data reduction.
This study is part of a project that has received funding from the European Research Council (ERC) under the European Union’s Horizon 2020 research and innovation programme under Grant agreement No. 949278 (Chemtrip) and Grant agreement No. 716155 (SACCRED). 
This work made use of Astropy:\footnote{http://www.astropy.org} a community-developed core Python package and an ecosystem of tools and resources for astronomy \citep{astropy:2013, astropy:2018, astropy:2022}.
\end{acknowledgements}

\bibliographystyle{aa}
\bibliography{Biblio} 
\begin{appendix}

\section{Additional tables and figures on observations}

\begin{table*}
\caption{List of calibrators.}
\begin{center}
\begin{tabular}{lccccc}
\hline
\hline
Project & Date & Configuration & Phase calibrators & Bandpass calibrator & Flux scale calibrator \\
\hline
W18AO & 15 February 2019 & A & 0507+179, 0446+112 & 3C84 & LkH$\alpha$ 101 \\
W18AO & 17 February 2019 & A & 0507+179, 0446+112 & 3C84 & LkH$\alpha$ 101, 2013+370 \\
W18AO & 27 February 2019 & C & 0507+179, 0446+112 & 3C84 & LkH$\alpha$ 101 \\
S16AE & 25 August 2016 & D & 0507+179, 0354+231 & 3C454.3 & MWC349 \\
S16AE & 31 August 2016 & D & 0507+179, 0354+231 & 3C454.3 & LkH$\alpha$ 101 \\
\hline
\end{tabular}
\end{center}
\label{tab:calibrators}
\end{table*}

\begin{table*}
\caption{Gaussian fitting parameters and fluxes of the water isotopolog lines.}
\begin{center}
\begin{tabular}{lccccc}
\hline
\hline
Molecule & Rest frequency [MHz] & $\varv_{\rm LSR}$ [km\,s$^{-1}$] & Peak intensity [K] & FWHM [km\,s$^{-1}$] & Flux [K.km\,s$^{-1}$] \\
\hline
\multicolumn{6}{c}{North source position (N)} \\
\hline
HDO & 225896.72 & 9.11 $\pm$ 0.06 & 5.22 $\pm$ 0.19 & 2.78 $\pm$ 0.15 & 15.46 $\pm$ 1.41 \\
 & & 5.71 $\pm$ 0.09 & 2.90 $\pm$ 0.23 & 1.96 $\pm$ 0.28 & 6.06 $\pm$ 1.34 \\
H$_2^{18}$O & 203407.52 & 9.71 $\pm$ 0.15 & 1.61 $\pm$ 0.21 & 2.25 $\pm$ 0.38 & 3.85 $\pm$ 1.16 \\
 & & 5.44 $\pm$ 0.14 & 2.22 $\pm$ 0.21 & 2.71 $\pm$ 0.48 & 6.39 $\pm$ 1.73 \\
\hline
\multicolumn{6}{c}{Middle position (M)} \\
\hline
HDO & 225896.72 & 5.73 $\pm$ 0.10 & 3.20 $\pm$ 0.25 & 2.01 $\pm$ 0.26 & 6.87 $\pm$ 1.43 \\
 & & 9.15 $\pm$ 0.09 & 4.06 $\pm$ 0.21 & 2.89 $\pm$ 0.23 & 12.48 $\pm$ 1.67 \\
H$_2^{18}$O & 203407.52 & 5.46 $\pm$ 0.09 & 2.06 $\pm$ 0.17 & 2.31 $\pm$ 0.26 & 5.07 $\pm$ 0.98 \\
 & & 9.54 $\pm$ 0.16 & 1.18 $\pm$ 0.17 & 2.29 $\pm$ 0.43 & 2.87 $\pm$ 0.96 \\
\hline
\multicolumn{6}{c}{N and M positions$^{(a)}$} \\
\hline
HDO & 143727.21 & & & & 1.07 $\pm$ 0.30 \\
\hline
\end{tabular}
\tablefoot{$^{(a)}$ The two components at 6 and 9 km s$^{-1}$ cannot be distinguished because of the poor spectral resolution of the data (1.95 MHz). The total integrated flux is consequently only given for this transition. A similar flux is obtained for the N and M positions because of the low spatial resolution of the data (2.5$\arcsec$ $\times$ 2.1$\arcsec$).}
\end{center}
\label{tab:gauss-param}
\end{table*}

\begin{figure*}[ht!]
    \centering
    \includegraphics[width=16cm]{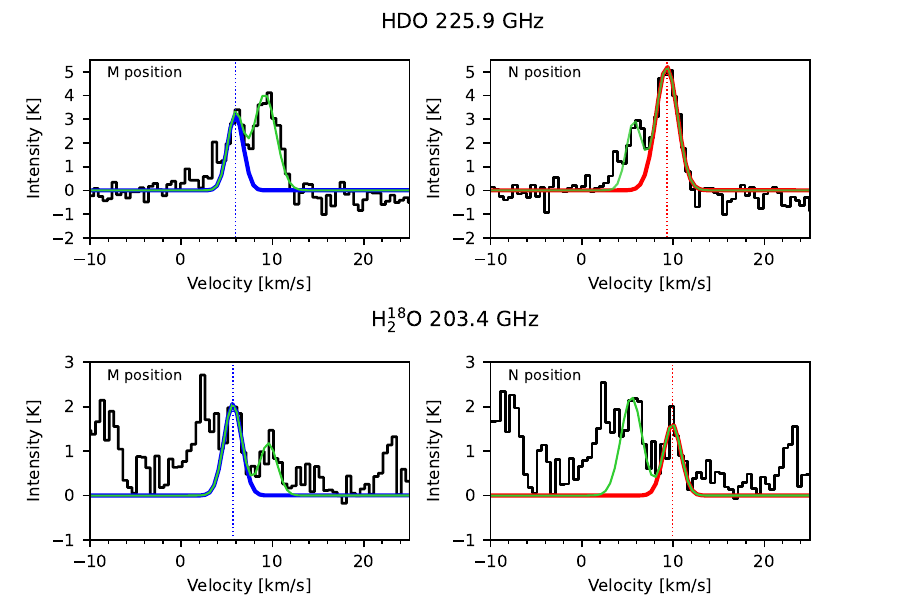}
    \caption{Gaussian line fits  using the CASSIS software. The upper panel shows the spectra of the HDO transition at 225.9 GHz, while the lower panel is for the H$_2^{18}$O transition at 203.4 GHz. The left  plots are for the M position with the fit of the component at 6 km\,s$^{-1}$ (blue), while the right  plots are for the N position with the fit of the component at 9 km\,s$^{-1}$ (red). The dotted lines indicate the central velocity of the respective Gaussian. The green line shows the corresponding double-Gaussian profiles (see Table \ref{tab:gauss-param}).}
    \label{fig:Gaussians}
\end{figure*}

Table \ref{tab:calibrators} lists the calibrators used for the different observations.
Gaussian fitting was performed on the spectra to extract the fluxes of the HDO and H$_2$$^{18}$O lines (see Figure \ref{fig:Gaussians}). A summary of the fit results is presented in Table \ref{tab:gauss-param}.
We also performed circular Gaussian fitting of the emission lines in the ($u$,$\varv$)-plane to estimate the size of the sources. The results of the fits are summarized in Table \ref{tab:uv-fits}.

\begin{table*}[ht!]
\caption{Results from the circular Gaussian fits performed in the ($u$,$\varv$)-plane.}
\begin{center}
\begin{tabular}{lccc}
\hline
\hline
Transition & R.A. & Dec & FWHP [$\arcsec$] \\
\hline
HDO 225 GHz & 04$^{\rm h}$31$^{\rm m}$34$\fs$158 (0.007) & +18$\degr$08$\arcmin$04.75 (0.01) & 0.45 (0.02) \\
6.7 - 12 km\,s$^{-1}$  & \\

\hline

H$_2^{18}$O 203 GHz & 04$^{\rm h}$31$^{\rm m}$34$\fs$155 (0.044) & +18$\degr$08$\arcmin$04.80 (0.05) & 0.45 (0.11) \\
6.7 - 11 km\,s$^{-1}$  &  \\

\hline

HDO 225 GHz & 04$^{\rm h}$31$^{\rm m}$34$\fs$162 (0.014) & +18$\degr$08$\arcmin$04.60 (0.02) & 0.35 (0.04) \\
3 - 7.3 km\,s$^{-1}$  &  \\
\hline

\end{tabular}
\tablefoot{The uncertainties are indicated in parentheses. }
\end{center}
\label{tab:uv-fits}
\end{table*}

\section{Exploration of the physical parameters and the LTE assumption}
\label{AppendixB}

\subsection{Impact of source size and excitation temperature on  HDO/H$_2$O ratio}
\label{AppendixB1}

\begin{figure*}[!ht]
    \centering
    \includegraphics[width=17cm]{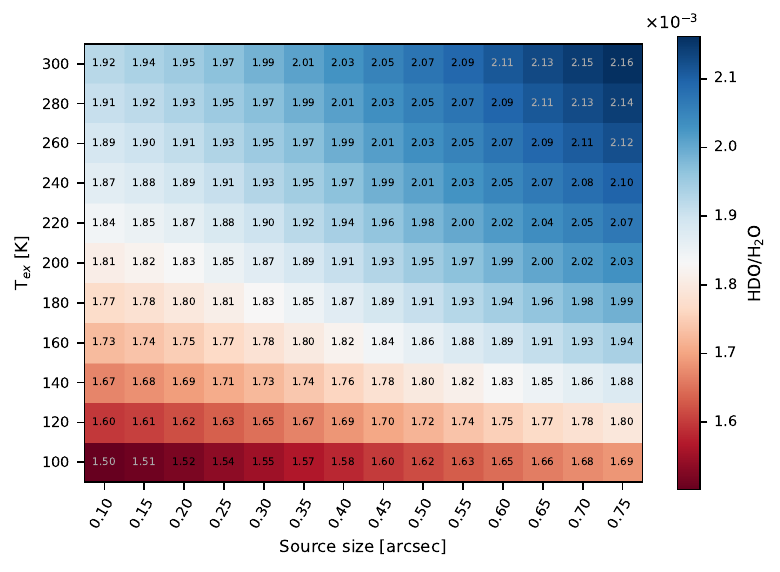}
    \caption{HDO/H$_2$O ratios ($\times$ 10$^{-3}$) obtained for the component at 9 km\,s$^{-1}$ (N) of L1551 IRS5 as a function of different excitation temperatures and sizes assuming optically thin LTE emission.}
    \label{fig:ratioevolves}
\end{figure*}

To investigate the impact of source size and excitation temperature on the HDO/H$_2$O ratio derived for the component at 9 km\,s$^{-1}$, we calculated the column densities needed to reproduce the fluxes of the HDO line at 225.9 GHz and the H$_2$$^{18}$O line at 203.4 GHz for a wide range of excitation temperatures (100--300\,K) and sizes (0.10--0.75$\arcsec$). 
The results of this method, which assumes LTE and optically thin lines, are presented in Figure \ref{fig:ratioevolves}. The derived HDO/H$_2$O ratio is not very sensitive to the excitation temperature and size as it ranges between 1.5\,$\times$\,10$^{-3}$ and 2.2\,$\times$\,10$^{-3}$. An analogous study done by \citet{Jorgensen2010} led to similar conclusions.

\begin{figure}[!ht]
    \resizebox{\hsize}{!}{
    \includegraphics[width=9cm]{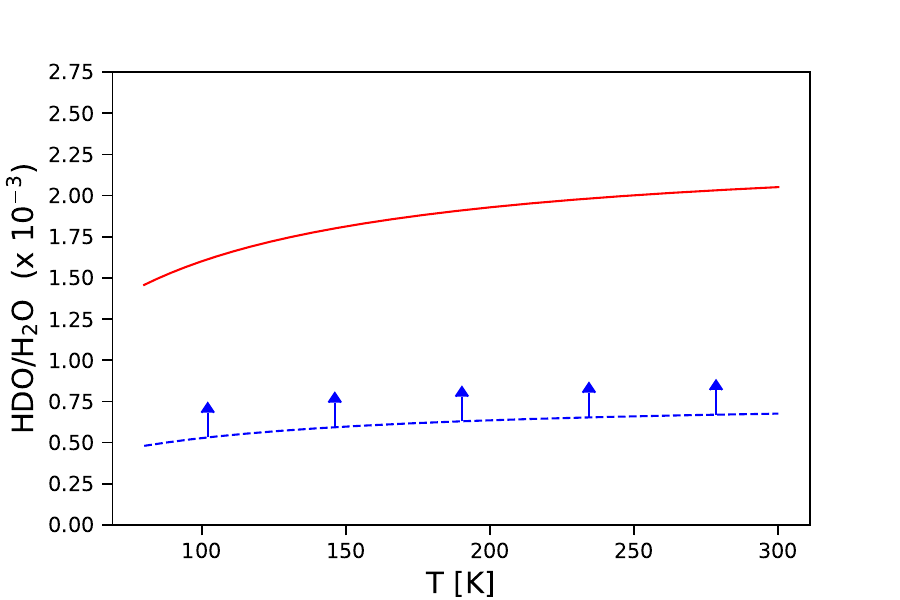}}
    \caption{Evolution of   HDO/H$_2$O ratio as a function of   excitation temperature assuming source sizes of 0.35$\arcsec$ and 0.45$\arcsec$ for the M and N positions, respectively. The solid red line represents the ratio found for N (component at 9 km\,s$^{-1}$), while the blue dashed line is for M (6 km\,s$^{-1}$). The arrows indicate   lower limits.}
    \label{fig:Tplotvar}
\end{figure}

Figure \ref{fig:Tplotvar} shows how the HDO/H$_2$O ratios of  both components vary as a function of the excitation temperature when using the source sizes derived with circular Gaussian fitting of the lines in the ($u$,$\varv$)-plane.

\subsection{Non-LTE modeling}
\label{AppendixB2}

\begin{figure}[!ht]
    \resizebox{\hsize}{!}{
    \includegraphics{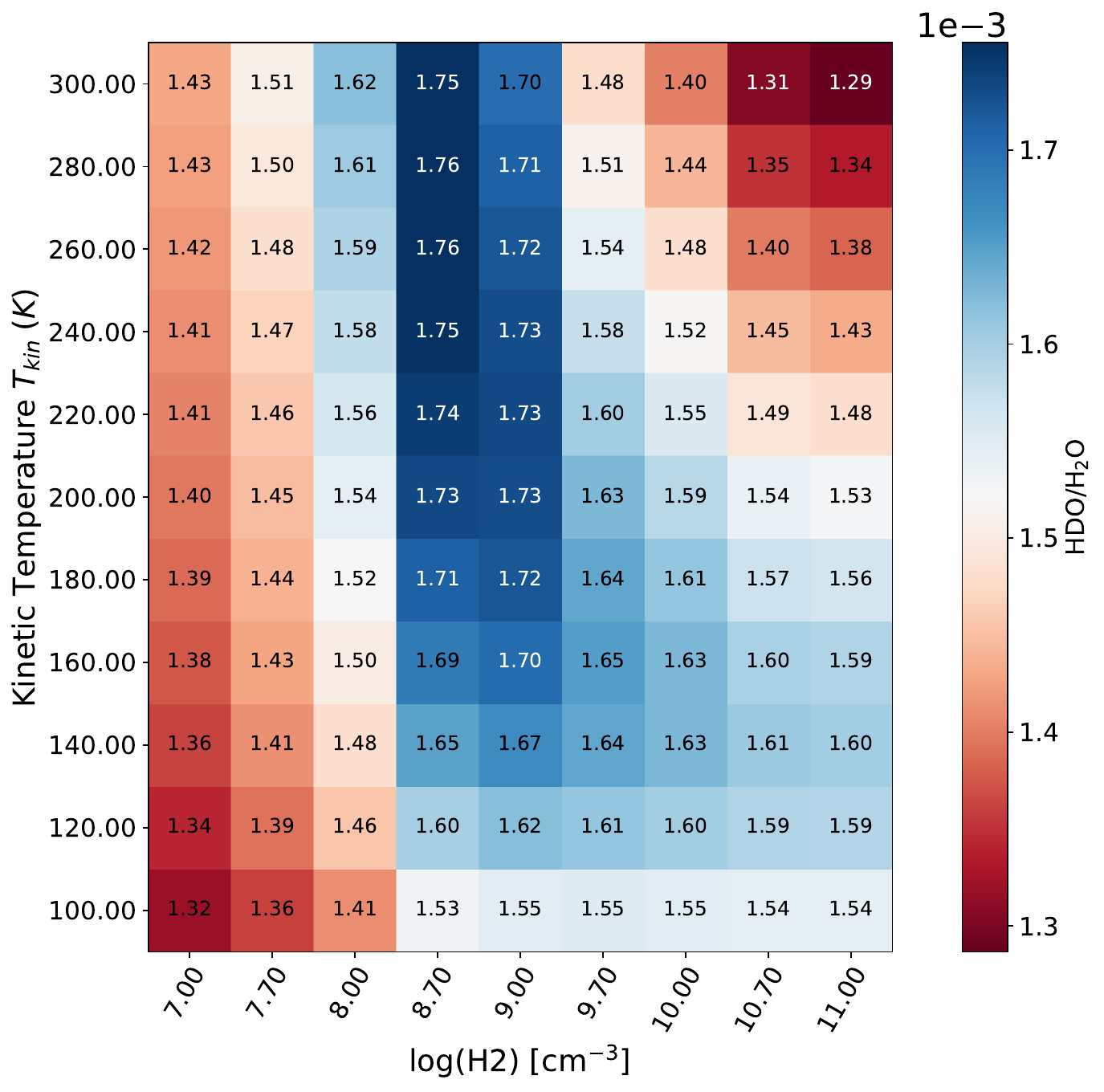}}
    \caption{HDO/H$_2$O ratio ($\times$ 10$^{-3}$) obtained for the component at 9 km\,s$^{-1}$ (N) of L1551 IRS5 with the non-LTE radiative transfer code RADEX as a function of the kinetic temperature and the H$_2$ density.}
    \label{fig:array-nonLTE}
\end{figure}

The LTE assumption has been tested and approved for the same transitions in other hot corinos \citep{Persson2014,Coutens2014}.
We carried out a similar test for L1551~IRS5 using the non-LTE radiative transfer code RADEX \citep{vanderTak2007} with the Large Velocity Gradient (LVG) expanding sphere method. 
We varied the temperature between 100 and 300 K, similarly to the LTE analysis. For the H$_2$ density, we considered a range from 10$^7$ to 10$^{11}$ cm$^{-3}$. With a non-LTE analysis of three HDO lines, \citet{Coutens2014} derived a H$_2$ density higher than 10$^{8}$ cm$^{-3}$ in the warm inner regions of the low-mass protostar NGC1333 IRAS2A. 
The results are shown in Figure \ref{fig:array-nonLTE}.
The HDO/H$_{2}$O ratio derived with RADEX, (1.3 - 1.8) $\times$ 10$^{-3}$ is in agreement with the LTE values within the uncertainties.

\subsection{Modeling of the HDO line at 143.7 GHz and constraints on the excitation temperatures}
\label{AppendixB3}

\begin{figure*}[!ht]
    \resizebox{\hsize}{!}{
    \includegraphics[width=15cm]{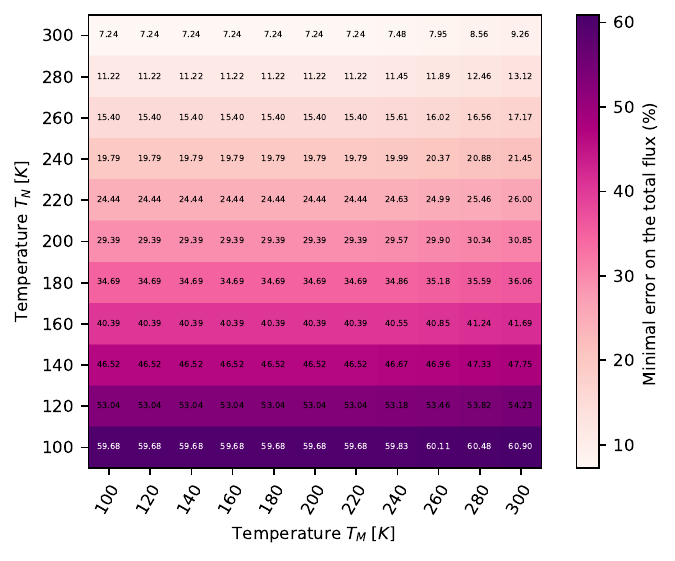}}
    \caption{Percentage errors between the predicted and observed fluxes of the HDO line at 143.7 GHz as a function of  excitation temperature for  positions M and N. Source sizes of 0.35$\arcsec$ and 0.45$\arcsec$ are assumed for the M and N positions, respectively.}
    \label{fig:HDO143}
\end{figure*}

The spectral resolution of the HDO transition at 143.7 GHz does not allow us to distinguish between the fluxes coming from the component at 6 km\,s$^{-1}$ and the one at 9 km\,s$^{-1}$. However, we can use the total flux to constrain the excitation temperatures for the two components. Assuming the source sizes obtained with the fits in the ($u$,$\varv$)-plane, we calculated the column densities of HDO needed to reproduce the two components at 225.9 GHz for various excitation temperatures. Then we predicted the expected flux at 143.7 GHz at 6 and 9 km\,s$^{-1}$, and summed the two fluxes before comparing them to the observed flux (see Table \ref{tab:gauss-param}). Figure \ref{fig:HDO143} shows that a good agreement ($<$ 20\%) is only found if the excitation temperature for the component at 9 km\,s$^{-1}$ is $\gtrsim$ 240 K. The excitation temperature is not constrained for the component at 6 km\,s$^{-1}$.

\section{Previous results on water deuteration}

Table \ref{tab:Jensen-ratios} summarizes the HDO/H$_2$O ratios determined in comets and protostars.

\begin{table}[!h]
\caption{Water deuterium fractionation ratios determined in comets and protostars.}
\begin{center}
\begin{tabular}{lcc}
\hline
\hline
Source name & HDO/H$_2$O  & Ref. \\
& ($\times$ 10$^{-4}$) & \\
\hline
\multicolumn{3}{c}{Oort Cloud Comets } \\
\hline
1/P Halley & 4.2 $\pm$ 0.6 & (1) \\
C/1996 B2 Hyatuake & 5.8 $\pm$ 2.0 & (2) \\
C/1995 O1 Hale-Bopp & 6.6 $\pm$ 1.6 & (3) \\
C/2007 B3 Lulin & < 11.2 & (4) \\
8P/Tuttle & 8.2 $\pm$ 2.9 & (5) \\
C/2009 P1 Garradd & 4.1 $\pm$ 0.4 & (6) \\
C/2002 T7 LINEAR & 5.0 $\pm$ 1.4 & (7) \\
153P Ikeya-Zhang & < (5.6 $\pm$ 0.6) & (8) \\
C/2012 F6 Lemmon & 13.0 $\pm$ 3.2 & (9) \\
CC/2014 Q2 Lovejoy & 2.8 $\pm$ 0.8 & (9) \\

\hline
\multicolumn{3}{c}{Jupiter Family Comets } \\
\hline
45P/Honda-Mrkos-Pajdu\u{s}áková & < 4.0 & (10) \\
103P Hartley 2 & 3.2 $\pm$ 0.5 & (11) \\
67/P Churyumov-Gerasimenko & 10.6 $\pm$ 1.4 & (12) \\
46P/Wirtanen & 3.2 $\pm$ 1.3 & (13) \\
\hline
\multicolumn{3}{c}{Clustered Class~0 protostars} \\
\hline
NGC1333 IRAS 4A-NW & 5.4 $\pm$ 1.5 & (14) \\
NGC1333 IRAS 2A & 7.4 $\pm$ 2.1 & (14) \\
NGC1333 IRAS 4B & 5.9 $\pm$ 2.6 & (14) \\
IRAS 16293-2422 & 9.2 $\pm$ 2.6 & (15) \\
\hline
\multicolumn{3}{c}{Isolated Class~0 protostars} \\
\hline
BHR71-IRS1 & 18 $\pm$ 4 & (16) \\
B335 & 17 $\pm$ 3 & (16) \\
L483 & 22 $\pm$ 4 & (16) \\
\hline
\multicolumn{3}{c}{Class~I protostars} \\
\hline
V883 Ori & 23 $\pm$ 6 & (17) \\
L1551 IRS5 -- $N, \varv_{\rm LSR}$ $\sim$ 9 km\,s$^{-1}$  & 21 $\pm$ 8 & (18) \\
L1551 IRS5 -- M, $\varv_{\rm LSR}$ $\sim$ 6 km\,s$^{-1}$  & $>$ (5 $\pm$ 2) & (18) \\
\hline
\end{tabular}
\tablefoot{This table is an updated version of Table A1 from \citet{Jensen2019}. A factor  2 error on the conversion of the D/H ratio onto the HDO/H$_2$O ratio has been corrected for the comets C/2012 F6 Lemmon and CC/2014 Q2 Lovejoy \citep{Biver2016}.}
\tablebib{ (1) \citet{Brown2012}; (2) \citet{Bockelee-Morvan1998}; (3) \citet{Meier1998}; (4) \citet{Gibb2012}; (5) \citet{Villanueva2009}; (6) \citet{Bockelee-Morvan2012}; (7) \citet{Hutsemekers2008}; (8) \citet{Biver2006}; (9) \citet{Biver2016}; (10) \citet{Lis2013}; (11) \citet{Hartogh2011}; (12) \citet{Altwegg2015}; (13) \citet{Lis2019}; (14) \citet{Persson2014}; (15) \citet{Persson2013}; (16) \citet{Jensen2019}; (17) \citet{Tobin2023}; (18) This work. }
\end{center}
\label{tab:Jensen-ratios}
\end{table}

\end{appendix}

\end{document}